\documentclass{article}

\usepackage{INTERSPEECH2019}
\usepackage{graphicx}
\usepackage{tabulary}
\usepackage{multirow}
\usepackage{makecell}
\usepackage{boldline}

\graphicspath{{./figs/}}
\DeclareGraphicsExtensions{.png}

\usepackage{cite}
\usepackage{bm}
\usepackage[varg]{txfonts}
\usepackage{mathptmx}	
\usepackage{times}
\usepackage{psfrag}

\def\tr{\mathop{\rm tr}}
\newcommand{\noise}{{\text{N}}}
\newcommand{\speech}{{\text{S}}}

\usepackage{amsmath}

\title{Meeting Transcription Using Virtual Microphone Arrays}
\name{Takuya Yoshioka, Zhuo Chen, Dimitrios Dimitriadis, William Hinthorn, Xuedong Huang \\Andreas Stolcke, Michael Zeng}
\address{Microsoft Technical Report MSR-TR-2019-11\\
Revised July 2019%
\thanks{This report is an expanded version of a conference paper \cite{Denmark:interspeech2019} with added results on NIST-RT data.}}
\email{\{tayoshio,zhuc,didimit,wihintho,xdh,anstolck,nzeng\}@microsoft.com}

\begin{document}

\maketitle
\begin{abstract}
We describe a system that generates speaker-annotated transcripts of meetings by using a virtual microphone array, a set of spatially distributed asynchronous recording devices such as laptops and mobile phones. The system is composed of continuous audio stream alignment, blind beamforming, speech recognition, speaker diarization using prior speaker information, and system combination. 
%
%
When utilizing seven input audio streams, our system achieves a word error rate (WER) of 22.3\% and comes within 3\% of the close-talking microphone WER on the non-overlapping speech segments. 
The speaker-attributed WER (SAWER) is 26.7\%.
The relative gains in SAWER over the single-device system are 14.8\%, 20.3\%, and 22.4\% for three, five, and seven microphones, respectively. The presented system achieves a 13.6\% diarization error rate when 10\% of the speech duration contains more than one speaker. The contribution of each component to the overall performance is also investigated, and we validate the system with
experiments on the NIST RT-07 conference meeting test set.
\end{abstract}

\section{Introduction}
Speaker-attributed automatic speech recognition (SA-ASR) of natural meetings has been one of the very challenging tasks since the early 2000s, when the NIST Rich Transcription Evaluation series \cite{FiscusEtAl:rt07} started.
Systems developed in the early days yielded high error rates, especially when distant microphones were used as input. 
However, with the rapid progress in conversational speech transcription~\cite{Saon17,XiongEtAl:icassp2018}j, far-field speech recognition~\cite{Yoshioka15b,Du16,Li17,Li18}, and speaker identification and diarization~\cite{Zhang18,Sell18}, 
realizing accurate meeting transcription from a distance seems to be within reach, especially when employing microphone arrays. 
In addition to the microphone array setups, single-microphone systems have also been evaluated.

The use of multiple unsynchronized audio streams, such as from mobile devices, adds complexity to the meeting setup and processing.
In return, we gain potentially better spatial coverage since the devices will tend to be distributed around the room and relatively near the speakers.  Also, in many use cases it will be natural for meeting participants to bring, and then repurpose their personal devices, in the service of better transcription quality.

On the other hand, while there are several pioneering studies~\cite{Araki18},
it is unclear what the best strategies are for consolidating multiple asynchronous audio streams and to what extent they work for natural meetings in online and offline setups. 

In this paper, we investigate a meeting transcription architecture based on asynchronous distant microphones by combining both front-end and back-end techniques. The resulting system performance is investigated on real-world meeting recordings. 
Our proposed system is designed to generate word recognition results in real time and
then provide improved speaker-attributed transcriptions with limited latency.

In addition to the end-to-end system analysis, we make the following specific contributions: we examine the idea of ``leave-one-out beamforming'' in the asynchronous multi-microphone setup. This method was proposed 
to benefit from both beamforming and system combination approaches but tested only with synchronized signals~\cite{Stolcke11}.
The computational cost required for calculating multiple beamformers can be reduced by taking advantage of the properties of spatial covariance matrices. 
We investigate a similar diversity-preserving strategy for acoustic model fusion.
Further, we describe three different system combination schemes that take account of both word recognition and speaker attribution.
Finally, we show results based on incremental ROVER that processes the ASR and diarization outputs with low latency.

The rest of the paper is organized as follows. 
Section \ref{sec: overview} describes the meeting transcription task that we consider in this paper and an overall architecture of our proposed system. 
Section \ref{sec: components} elaborates on individual system components with emphasis on unique aspects of our work. 
Section \ref{sec: results} reports experimental results, followed by concluding remarks in Section \ref{sec: conclusion}.

\begin{figure*}[t]
\centering
\includegraphics[scale=0.67]{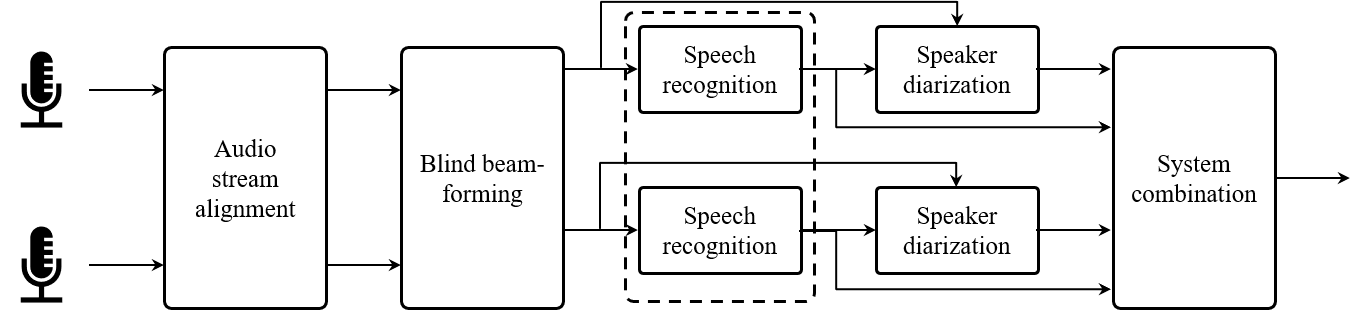}
\caption{Processing flow diagram of proposed meeting transcription system with asynchronous multi-microphone input. ASR may be performed jointly across channels as implied by the dotted box.} 
\label{fig: overall}
\end{figure*}

\section{Task and System Overview}
\label{sec: overview}
We record a meeting with $M$ audio capturing devices, 
such as cell phones, tablets, and laptops. 
The devices can be randomly placed  in a conference room. 
The acoustic signal picked up by each device is transmitted to a common server. 
The server then generates a speaker-attributed transcription of the meeting conversation 
in real time as it receives the signals from the devices. 
In this paper, we assume that all meeting attendees have  enrolled in the system and have provided their voice-prints for speaker identification. 

Figure~\ref{fig: overall} shows the processing flow of the proposed system. 
The input signals received by the server are misaligned for various reasons such as clock drift on each recording device, differences in on-device signal processing, packet generation, and signal transmission channels. 
As in Fig.~\ref{fig: overall}, the audio stream alignment module constantly corrects the inter-channel signal misalignments. 
This is followed by a beamforming module, which receives the $M$ time-aligned audio signals and yields $N$ enhanced signals. In this paper, we deal with the case where $M = N$ while this is not a requirement. 
Each enhanced signal is fed to a speech recognition module to produce a
real-time transcription as well as n-best recognition hypotheses with word-level time marks. 
The diarization module then generates a speaker label sequence for each of the segments detected by the ASR decoder by utilizing both word-level time marks and speaker embeddings extracted from the enhanced audio. 
Eventually, the speaker labels and word hypotheses are processed in the system combination module to yield a final speaker-attributed word transcription. 

We have settled on this architecture based on several considerations:
it supports both beamforming and later-stage system combination approaches, which are found to be beneficial together \cite{Stolcke11}. 
Also, we perform diarization after speech recognition, unlike in many previous systems. Since diarization typically has a longer algorithmic delay this allows preliminary recognition results to be displayed in real time.
Finally, system combination, coming last, is designed to merge and benefit both word recognition and speaker attribution.

\section{System Components}
\label{sec: components}

\subsection{Audio Stream Alignment}

The audio stream alignment module picks one of the input streams as a reference and aligns each of the other signals to the reference signal. 
When aligning a signal to the reference, the time lag between the two signals is first detected and then  the non-reference signal is adjusted accordingly.
For this purpose, we have two variable-length first-in-first-out buffers for each stream: one for time lag detection, one for output generation. 
After a few seconds (2\,s in our experiments), we extract as many samples from the output buffer as those pushed to the buffer. These samples are given to the downstream modules. 

At $T$-second intervals ($T=30$\,s in our experiments), we calculate the cross-correlation coefficients between the two signals stored in the time-lag detection buffer and pick the sample lag $L$ that maximizes the cross-correlation value. We decimate the samples in the non-reference stream output buffer by $|L|$ if $L > 0$. Otherwise, we increase the number of samples by $|L|$ by resampling. The time-lag detection buffer is then refreshed. 

Two sources of misalignment between the signals can be distinguished, the ``global'' and the ``local'' ones. The global lag is due to the beginning of the recordings and it appears fixed. The local lag is due to clock drifting, channel delays, windows, etc. The local lag is variable and time-varying, thus requires continuous synchronization.
At the beginning of the alignment processing, we may calculate the cross-correlation more frequently (e.g., every 1\,s) until we find 
a significant peak in the cross-correlation sequence. 
This `global' time lag can also be used to adjust the output wait time. In an online client-server setting, the global time lag is small. 
When we apply the system to offline independent recordings, the global time lag can be in the order of minutes. In this case, we may use a sliding window to first obtain an approximate estimate of the global time lag and then fine-tune the local estimates by using the sample-level cross-correlation as described above.

\subsection{Blind beamforming}

For beamforming, we adopt a mask-based blind processing approach~\cite{Heymann16,Higuchi16}. 
This approach was shown to perform as well as carefully designed beamformers that utilize array geometry information~\cite{Boeddeker18}.

\vspace{1em}
\noindent{\bf Mask-based blind beamforming.}
Assuming $M$ microphones to be available, 
an enhanced short time Fourier transform (STFT) coefficients can be computed as the inner product of 
$M$-dimensional beamformer coefficient vector $\bm{w}_f$ and input multi-channel STFT coefficient vector $\bm{y}_{ft}$, 
where subscripts $f$ and $t$ denote frequency bin and time frame indices, respectively.
In one formulation, the beamformer coefficients are estimated with a minimum variance distortionless response (MVDR) principle as
\begin{align}
\bm{w}_f = \frac{\bm{\Phi}_{\noise, f}^{-1} \bm{\Phi}_{\speech, f} \bm{r}}
{\tr \bigl(\bm{\Phi}_{\noise, f}^{-1} \bm{\Phi}_{\speech, f}\bigr)},
\label{eq: beamform}
\end{align}
where $\bm{\Phi}_{i}$ is the spatial covariance matrix ($i=\speech$ for speech; $i=\noise$ for noise) where $\bm{r}$  a one-hot unit vector with $1$  at the position of the  reference microphones, which may be chosen based on a maximum signal-to-noise ratio (SNR) principle~\cite{Erdogan16}.  
The speech and noise spatial covariance matrices are estimated using spectral masks, i.e.,  
\begin{align}
\bm{\Phi}_i = \frac{1}{\sum_{t \in \mathbb{T}} M_{i, ft}} \sum_{t \in \mathbb{T}} M_{i, ft} \bm{y}_{ft} \bm{y}_{ft}^H, 
\end{align}
where 
$\mathbb{T}$ is a segment over which the beamformer is estimated, and 
$M_{i, ft}$ is the spectral mask for time-frequency point $(f, t)$. 

In our experiments, a neural network trained to minimize the mean squared error between clean and enhanced log-Mel features was used~\cite{Boeddeker18}. 
The spectral masks were estimated for every 1\,s-batch. The beamformer coefficients are also updated accordingly.


\vspace{1em}
\noindent{\bf Strategies for generating multiple different outputs.}
\label{subsec: leave_one_out}
System combination relies on errors being partly uncorrelated among inputs.
For this reason, \cite{Stolcke11} suggested manipulating early-fusion approaches to keep the outputs as decorrelated as possible, specifically using a leave-one-out approach to beamforming.
Two such schemes are investigated in this work. 

In the first scheme, called the all-channel approach, 
we rotate the 1's position in unit vector $\bm{r}$ from the 
first element to the last to create different beamformer coefficient vectors 
based on Eqn.~\eqref{eq: beamform}. 
A potential drawback of this approach is that 
the beamformer outputs might not retain enough diversity among different channels because they are still based on the same input signals.

The second, ``leave-one-out'' (LOO) scheme forms an acoustic beam by using 
$M-1$ channels while varying the left-out microphone in a round-robin manner. 
This scheme requires $M$ different $(M-1)$-dimensional noise spatial covariance matrices to be inverted 
in order to calculate $M$ beamformers based on Eqn.~\eqref{eq: beamform}. 
It can be shown that all the $M$ inverse spatial covariance matrices of size $M-1$ can be derived from a shared $M$-dimensional inverse spatial covariance matrix by utilizing
the matrix inversion properties of block and permutation matrices.  
Therefore, both two schemes can be run with similar computational cost.

\subsection{Speech recognition}

The speech recognition module converts an incoming audio signal to 
an n-best list with word-level time marks. 
In the experiments reported later, we used a conventional hybrid ASR system, consisting of a latency-controlled bidirectional long short-term memory (LSTM) acoustic  model (AM) \cite{Xue17} and a weighted finite state transducer decoder.
Our AM was trained on 33K hours of in-house audio data, including close-talking, distant-microphone, and artificially noise-corrupted speech. Decoding was performed with a trigram language model (LM).
Whenever a silence segment longer than 300\,ms was detected, the decoder generated an n-best list, which was rescored with both a 5-gram trained on 100B words and an LSTM-LM.
The latter used two 2048-unit recurrent layers and was trained on 2B words.

\subsection{Speaker diarization with prior information}
\label{subsec:Diarization}
Given a speech region detected by the speech recognition module,
speaker diarization assigns a person label to each word in the 
top recognition hypothesis. 
We adopt an approach consisting of three steps: d-vector generation, segmentation, and speaker identification. 
With our decoder configuration, each incoming speech region typically contains up to 20 words. 

The d-vector generation step calculates speaker embeddings~\cite{Variani14} for every fixed time interval (320\,ms in our system). 
We trained a ResNet-style embedding extraction network~\cite{He16} on the VoxCeleb corpus~\cite{VoxCeleb} to generate 128-dimensional d-vectors.

The speaker segmentation step decomposes the received word sequence into speaker-homogeneous subsegments. This is performed with an agglomerative clustering approach~\cite{Gauvain98,Tranter06} by using the d-vectors as observed samples. 
Initially, every single word comprises a unique subsegment. 
For every neighboring subsegment pair, the degree of proximity between the two subsegments is estimated in the embedding space. The closest pair is then merged to form a new subsegment. The proximity is defined as the cosine similarity between the 
mean d-vectors. This process is repeated until the cosine similarity drops below a threshold (0.15 in our experiments).

Finally, a speaker label is assigned to each subsegment. 
In this paper, we assume that a list of meeting attendees is available. 
For each subsegment, a segment-level embedding is computed by averaging the d-vectors over the subsegment. 
Likewise, the embedding of each speaker is pre-computed from enrollment audio samples, which were around 30\,s long. 
The speaker label that gives the highest cosine similarity to the subsegment embedding is selected.

\subsection{System combination}


System combination consolidates the multiple speaker-attributed ASR results to produce 
a final transcription result. 
ROVER~\cite{Fiscus97} and confusion network combination (CNC)~\cite{StolckeEtAl:nist2000,EvermannWoodland:nist2000}
are two popular system combination approaches. 
The goal of this step is to combine evidence from all channels, after beamforming, for both word and speaker recognition. 
As discussed in Section~\ref{subsec:Diarization}, a speaker label is assigned to every word based on the acoustics of the available audio streams. For purposes of ROVER, the speaker identities are encoded as audio channel numbers. 
Then,  they  are  submitted  to  the  NIST
ROVER algorithm \cite{Fiscus97} along with the word hypotheses, which
combines them by aligning words based on dynamic programming and their time marks
and  extracting  the  words  with  the  highest  vote  count.
We have modified the interface to the ROVER algorithm in such a way that this process can be invoked online, as new speaker-attributed word hypotheses become available from the diarization module, by using a sliding window shared across streams. Due to  misalignment between different decoder outputs, some words may appear twice. We run a simple filter removing the duplicates.

For CNC-based system combination, we devised an alternative algorithm that currently operates in batch mode.
On each channel, for each speech segment, the decoder generates n-best lists, which are aligned into confusion networks (CNs). The speaker recognition output from each channel is also encoded as a CN, using special tags for the speaker identities, interspersed with 1-best word hypotheses.
We modified the CN algorithms in SRILM \cite{srilm:icslp2002} to support aligning word and speaker CNs, and augmented the usual minimal edit distance objective function with a time-misalignment penalty.  The end result of the modified CNC is that n-best word hypotheses from all channels are merged with the speaker information, and the speakers and words with highest combined posteriors can be decoded jointly.

\subsection{Acoustic model combination}

In addition to the channel-fusion approaches described above, i.e., beamforming and system combination, it is also possible to combine frame-level senone posterior probabilities from multiple streams before ASR decoding~\cite{TibrewalaHermansky:icassp97}. 
While this approach is not integrated into the end-to-end system yet, we have investigated the effectiveness of senone-level AM fusion, with strategies aimed at increasing the diversity of the output results for later processing with ROVER or CNC.

The baseline results (first row) in Table~\ref{tab:AM_Comb} use senone posteriors from a single channel, produced by the AM and used as input to the decoder. 
Next, the sum and max of senone posteriors across channels are investigated.
This results in a single word hypothesis stream, with ROVER/CNC combining speaker hypotheses only.
Similar to the leave-one-out strategy for beamforming, we can preserve diversity by sampling from the channels, followed by hypothesis combination. 
In the last two rows of Table~\ref{tab:AM_Comb}, 
we present results with 6-out-7 senone fusion (resulting in 7 different senone subsets), and 3-out-7 with 35 outputs.  In the latter case, we sample 7 of the 35 possible outputs to reduce computation.  Either way, the 7 resulting decoding outputs are routed to system combination as before.

\begin{table}[tb]
\centering
\caption{AM combination. Results for one particular meeting. Numbers should not be compared with 
those of other tables.}
\vspace{-1em}
\begin{tabulary}{\linewidth}{L|c|c|c|c}
\Xhline{3\arrayrulewidth}
                         & \multicolumn{2}{c|}{ROVER} & \multicolumn{2}{c}{CNC} \\
                        & \footnotesize\%WER   & \footnotesize\%SAWER   & \footnotesize\%WER   & \footnotesize\%SAWER \\ \hline
Baseline                & 25.9    & 28.2      & 25.9        & 28.2      \\ 
Sum                     & 25.4    & 28.4      & 24.7    & 28.2  \\ 
Max                     & 22.5    & 27.5      & 22.5    & 27.2  \\ 
Max 6 of 7            & 23.8    & 27.2      & 22.1    & 26.7  \\ 
Max 3 of 7            & 24.2    & 26.8      & 22.3    & 26.9  \\ \hline
\end{tabulary}
\label{tab:AM_Comb}
\end{table}

\section{Experiments and Results}
\label{sec: results}

\subsection{Data and metrics}

We conducted a series of experiments to analyze the performance of 
the system described so far. 
We recorded five internal meetings; three meetings were recorded with seven independent consumer devices, four of which were iOS devices and three based on Android. All devices were different products. 
The other two meetings were recorded with a seven-channel circular microphone array. For these meetings, we did not make use of the fact that the signals were synchronous and let the signals through the entire pipeline including the audio stream alignment module. 
Those meetings took place in several different rooms and lasted for 30 minutes to one hour each, with three to eleven participants per meeting.
The meetings were neither scripted nor staged;
the participants conducted normal work discussions and were familiar with each other.
Partly as a result, about 10\% of all speech occurred in overlap with at least one other speaker.  Reference transcriptions were created by professional transcribers based on both close-talking and far-field recordings.

To test for generalization to different recording conditions, 
we also ran our system on meeting recordings from the NIST 2007 Rich Transcription (RT-07) evaluation \cite{FiscusEtAl:rt07}.
This RT-07 ``conference meeting'' test set consists of 8 meetings from four different recording sites, of varying lengths and with the number of microphones ranging from 3 to 16.
Each meeting has from four to six participants, with 31 distinct speakers in total.
Transcription accuracy is evaluated on a 22-minute excerpt from each meeting.

The system outputs were scored with NIST's scoring toolkit~\cite{Fiscus06} to calculate both standard, speaker-agnostic word error rates (WERs) and speaker-attributed WERs (SAWERs).
For the latter, a word is counted as correct only if both the word label and its speaker are identified correctly.
Note that these metrics count overlapped speech as any other.
Since our system, at present, does not attempt to separate overlapping speech we thus have a floor on the error rate of about 10\% (on the internal meeting data).

\begin{table}[tb]
\centering
\caption{WERs and SAWERs using seven microphones.}
\label{tab: results}
\vspace{-1em}
\begin{tabular}{ll | cc}
\Xhline{3\arrayrulewidth}
Sys. Comb. & Beamforming & \%WER & \%SAWER \\ \hline
None        & None          & 27.0 & 34.4 \\
(real time) & All channels  & 24.8 & 30.8 \\
            & Leave one out & 24.9 & 30.9 \\ \hline
ROVER       & None          & 25.3 & 28.5 \\ 
(online)    & All channels  & 24.2 & 27.4 \\ 
            & Leave one out & 24.2 & 27.2 \\ \hline
CNC         & None          & 22.8 & 27.7 \\
(offline)   & All channels  & 22.5 & 26.9 \\
            & Leave one out & 22.3 & 26.7 \\ \hline
\multicolumn{2}{l|}{IHM + reference diarization} & 14.4 & 14.4 \\
\hline
\end{tabular}
\end{table}

\subsection{Speech transcription accuracy on internal meetings}

Table~\ref{tab: results} shows the results for various system configuration, on
the meetings we recorded internally.
For the systems that do not perform any form of system combination, 
seven different results were obtained, each corresponding to a different one of the microphones, and 
the averages are reported in the table.
As a best case condition, and to calibrate the difficulty of the distant-microphone task, the final table row gives results for individual head-mounted microphones (IHM), with reference speaker segmentation.

The best system, combining beamforming and CNC, achieved substantial improvement 
over the single microphone system. The WER and SAWER relative gains were 
17.4\% and 22.4\%, respectively.
Relative to the IHM scenario as a floor, WER and SAWER were reduced by 37\%
and 39\%, respectively.

We can see that both beamforming and system combination (either with 
ROVER or CNC) contributed to the final performance, even though both steps combined information across channels. 
CNC provided the largest performance gain. While beamforming yielded a smaller gain, it is more easily used for real-time applications. 
The leave-one-out scheme provided slightly larger gains than 
the all-channel beamforming when combined with system combination, especially CNC, confirming our rationale in Section~\ref{subsec: leave_one_out}.


\begin{table}[tb]
\centering
\caption{Impact of number of microphones for system based on leave-one-out beamforming and CNC.}
\label{tab: nmics}
\vspace{-1em}
\begin{tabular}{lcccc}
\Xhline{3\arrayrulewidth}
No.~of microphones & 1 & 3 & 5 & 7 \\ \hline
\%WER & 27.0 & 24.0 & 22.7 & 22.3 \\
\%SAWER & 34.4 & 29.3 & 27.4 & 26.7 \\ 
\hline
\end{tabular}
\vspace{-1em}
\end{table}

Table \ref{tab: nmics} shows the WERs and SAWERs for different numbers of microphones.
There is a clear correlation between the number of microphones and the amount of improvement over the single channel system. Even with only three microphones, our system yielded relative gains of 11.1\% and 14.8\% in WER and SAWER, respectively.


\begin{table}[tb]
\centering
\caption{Speaker-independent WERs for non-overlapped segments. SDM: single distant microphone. BF: beamforming}
\label{tab:overlap}
\vspace{-1em}
\begin{tabular}{l|ccc|c}
\Xhline{3\arrayrulewidth}
System & SDM & BF & BF + CNC & IHM\\ \hline
\%WER & 20.6 & 18.1 & 16.2 & 13.2 \\\hline
\end{tabular}
\end{table}

To assess the speech recognition accuracy when a single person is speaking, 
we scored the results only against segments that did not contain any forms of overlap.%
\footnote{This was done by using NIST's asclite with the ``-overlap-limit 1'' option.}
Note that this discarded 58\% of the words. 
The results are shown in Table~\ref{tab:overlap}.
By comparing the numbers with the results of Table \ref{tab: results}, we can see that 
the system produced around 25\% more accurate transcriptions for the non-overlapped segments.
For the full system, the WER on non-overlapped speech is only 3.0\% worse than with close-talking microphones. 
Considering that the overlaps make up about 10\% of the speech duration, this result shows that segments including overlaps are more affected by the speaker-microphone distance. 


\subsection{Speech transcription accuracy on RT-07 meetings}

While these meeting recordings in the NIST RT-07 set are already synchronized,
we ran the front-end processing unchanged, and simply removed the word
deduplication step in the final hypothesis processing.
We evaluated the system in all three NIST evaluation conditions: single distant microphone (SDM), multiple distance microphones (MDM), and close-talking microphones (IHM).
SDM uses a single microphone designated by NIST, supposedly centrally located.
SDM and MDM used automatic speech detection, whereas IHM was run using the reference 
segmentation, since this task otherwise requires cross-talk suppression and we did not want to 
confound the results.%
\footnote{Teams participating in the original RT-07 evaluation typically found that automatic IHM segmentation gave WERs about 1-3\% higher than with reference segments \cite{StolckeEtAl:nist2007}.
Note that error rates reported here are much lower than in the original evaluation, reflecting general advances in speech recognition.
Also, the training data for the original evaluation was limited to shared, publicly available corpora, while our system had no such restriction.}
Another evaluation variable is the degree of overlap allowed in the scored segments.
Here we scored the outputs both on the non-overlapped segments and those with up to four overlapping speakers.%
\footnote{We stopped at four because the run-time for scoring grows exponentially with the allowed degree of overlap.
With this limit, only 2.3\% of all transcribed words are excluded.}

\begin{table}[tb]
    \centering
    \caption{Word error rate on NIST RT-07 meetings.
    The three MDM versions use no beamforming, beamforming on all microphones, and the leave-one-one approach, respectively.}
    \label{tab:rt07wer}
    \vspace{-1em}
    \begin{tabular}{l|c|c}
        \Xhline{3\arrayrulewidth}
        Evaluation condition      & No overlap  & Overlap $\leq 4$  \\
        \hline
        SDM                 & 16.7              & 28.2      \\
        MDM: CNC            & 15.5              & 26.2      \\
        MDM: All-mic-BF + CNC   & 14.8              & 26.3      \\
        MDM: LOO-BF + CNC   & 14.6              & 26.0      \\
        IHM                 & 12.3              & 15.9      \\
        \hline
    \end{tabular}
\end{table}

Table~\ref{tab:rt07wer} summarizes the results.
First, note that IHM error rates are very similar to those on our internal meetings (12-13\% for non-overlapping speech), showing that the intrinsic speech recognition difficulty is similar.
The SDM WER (28.2\%) is also similar to the average single-microphone WER in the earlier tests (27.0\%).
However, the relative WER reduction from multi-microphone processing (using both CNC and LOO-beamforming), is only 7.8\%.
This is much less than the 17.4\% reduction for the internal meeting set (comparing the first line in Table~\ref{tab: results} with the next-to-last line).
One reason could be that 
the microphones in most meetings of this test set were located so closely that 
the extra microphones yield less additional information,
compared with with the distributed, multiple-device setup used in our internal meetings. 

Although the performance gains from the MDM processing were smaller, 
the algorithms used in our system still each give substantial gains, especially on non-overlapped speech segments.
Beamforming in conjunction with CNC is 4.5\% better than CNC alone,
and leave-one-out processing improves the relative gain to 5.8\%.
Overall, the WER for non-overlapped speech that is achieved with multiple microphones is 12.6\% relative lower than for SDM, and only 2.3\% higher than with close-talking microphones.
This gap between MDM and IHM is again similar to the 3.0\% gap found for the internal meetings.

For a further examination of the effect of number of microphones on the different system versions, we chose the 5 (out of 8) RT-07 meetings with six or more microphones.
We then ran recognition with subsets of 1, 2, \ldots, 6 microphones, taking care to always include the SDM channel, and making sure that the subsets were nested (i.e., never removing any microphones).

\begin{figure}[t]
\hspace{-0.5\columnsep}%
\includegraphics[scale=0.67]{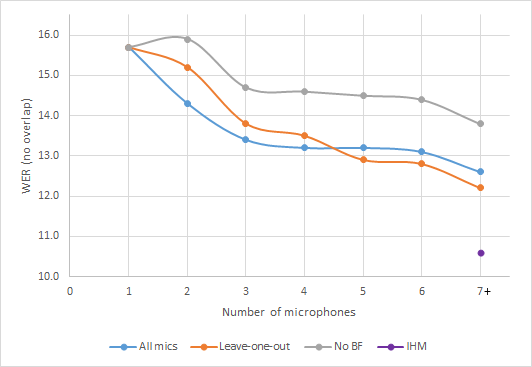}
\caption{Effect of varying number of microphones for different processing strategies (WERs on NIST RT-06 5-meeting subset)} 
\label{fig: rt07varying-microphones}
\end{figure}

Figure~\ref{fig: rt07varying-microphones} plots the WERs on non-overlapping segments of the 5-meeting subset,
for CNC without beamforming (No BF), all-microphone beamforming, and leave-one-out beamforming.
The right-most datapoint (``7+'') on each line refers to processing with all available microphones (up to 16 for some meetings).
The IHM error rate is shown as a single datapoint (at lower right).

A few observations are noteworthy.
First, without beamforming (CNC only) we see a degradation going from one to two microphones.
This is not surprising if the second microphone has much worse recognition than the first (SDM) one, since CNC does not work reliably if the input systems are of very different quality (unless that fact is known in advance and the systems can be weighted appropriately). 
Beamforming has the effect of creating audio streams of similar quality, thereby avoiding this problem for CNC.

Second, the leave-one-out strategy is not advantageous with four or fewer microphones.
This can be understood from the fact that removing a microphone from the beam in itself will make the results worse, and is only a win because the effectiveness of CNC is improved.
If very few microphones are available to begin with, the degradation due to the first effect can 
be larger than the gain from the second.
Therefore, when used in conjunction with system combination,
the beamforming strategy should be chosen depending on the number of available microphones.

\subsection{Speaker diarization accuracy}

\begin{table}[tb]
    \centering
    \caption{Diarization error on speaker-attributed ASR output. The percentage of overlapped speech is 10.0\%, and accounts for most of the missed speech.}
    \label{tab:der}
    \vspace{-1em}
    \begin{tabular}{l|ccc|c}
        \Xhline{3\arrayrulewidth}
                    & Misses & FAlarms & SpkrErr & DER \\
        \hline
         Avg. by channel & 10.5 & 3.3 & 1.8 & 15.6 \\
         CNC output & 10.2 & 2.4 & 1.0 & 13.6 \\
        \hline
    \end{tabular}
\end{table}

Since the speaker attribution algorithm at present relies on speaker enrollment, we evaluated its accuracy on internal meetings only.
We took the speaker-attributed recognition output, added 0.5\,s of extra duration at the margins of contiguous output from the same speaker, and evaluated the result according to the NIST ``Who spoke when'' task \cite{Tranter06}.
Note that our task is not speaker-agnostic diarization, but recognizing the known speakers.
Also, we are not trying to recognize overlapping speakers, so about 10\% of speech is missed, thus putting a floor on the missed speech and overall diarization error rate (DER).

Table~\ref{tab:der} gives the speaker diarization error of the system, by channel and for the combined output.  
The false alarm rate is quite low since the recognizer acts as a very conservative speech detection engine.
Similar to word recognition, CNC reduces the speaker error (44\% relative) by pooling speaker label posterior probabilities across all channels.

\section{Conclusion}
\label{sec: conclusion}
We studied a meeting transcription architecture for asynchronous distant microphones, combining front-end and back-end techniques, and evaluated it on real meeting recordings. We found that both front-end (blind beamforming) and back-end (model or system combination) algorithms improve word error, speaker-attributed word error, and diarization error metrics. 
Both beamforming and senone posterior fusion can be made more effective in conjunction with system combination by using  leave-one-out techniques. System combination was generalized such that it benefits both word and speaker hypotheses.
On non-overlapped speech, the error rate is only 3.0\% absolute worse than with close-talking microphones. 
We found broadly consistent results on NIST meeting evaluation data, with 2.3\% absolute WER difference between distant and close-talking microphones, for non-overlapped speech.
In summary, our study shows the effectiveness of multiple asynchronous microphones for meeting transcription in real-world scenarios. A major remaining challenge is recognition of overlapped speech~\cite{Yoshioka18b}.

\bibliographystyle{IEEEtran}

\bibliography{Interspeech2019}

\end{document}